# A Finely Segmented Semi-Monolithic Detector tailored for High-Resolution PET

Yannick Kuhl[1], Florian Mueller[1], Stephan Naunheim[1], Matthias Bovelett[1], Janko Lambertus[1], David Schug[1,2], Bjoern Weissler[1,2], Eike Gegenmantel[2], Pierre Gebhardt[1], Volkmar Schulz[1,2,3]

**Abstract**

**Background:** Preclinical research and organ-dedicated applications use and require high (spatial-)resolution positron emission tomography (PET) detectors to (early) visualize small structures and understand biological processes at a finer level of detail. Researchers seeking to improve detector and image spatial resolution have explored various high-resolution detector designs. Current commercial systems often employ finely pixelated or monolithic scintillators, each with its limitations.

**Purpose:** We present a semi-monolithic detector, tailored for high-resolution PET applications, and merging concepts of monolithic and pixelated crystals. The detector features slabs measuring $(24 \times 10 \times 1)$ mm³, coupled to a $12 \times 12$ readout channel photosensor with 4 mm pitch. The slabs are grouped in two arrays of 44 slabs each to achieve a higher optical photon density despite the fine segmentation.

**Methods:** We employ a fan beam collimator for fast calibration to train machine-learning-based positioning models for all three dimensions, including slab identification and depth-of-interaction (DOI), utilizing gradient tree boosting (GTB). The data for all dimensions was acquired in less than 2 hours. Energy calculation was based on a position-dependent energy calibration. Using an analytical timing calibration, time skews were corrected for coincidence timing resolution (CTR) estimation.

**Results:** Leveraging machine-learning-based calibration in all three dimensions, we achieved high detector spatial resolution: down to 1.18 mm full width at half maximum (FWHM) detector spatial resolution and 0.75 mm mean absolute error (MAE) in the planar-monolithic direction, 2.14 mm FWHM and 1.03 mm MAE for depth-of-interaction (DOI) at an energy window of (435-585) keV. Correct slab interaction identification in planar-segmented direction exceeded 80%, alongside an energy resolution of 13.8% and a CTR of 450 ps FWHM.

**Conclusions:** The introduced finely segmented, high-resolution slab detector demonstrates appealing performance characteristics suitable for high-resolution PET applications. The current benchtop-based detector calibration routine allows these detectors to be used in PET systems.

## I. INTRODUCTION

High-resolution positron emission tomography (PET) with high-resolution PET detectors plays a crucial role in preclinical research [1]–[4] and for organ-dedicated applications [5], [6] by enabling the analysis of physiological processes in small structures. Furthermore, heterogeneous detector groups can add value to PET systems [7], e.g., by locally improving the spatial resolution of the PET image with dedicated detector modules [8]–[13]. The HD-MetaPET project takes this idea a step further and aims to incorporate locally flexible high-resolution detectors into a long axial field-of-view (LAFOV)

PET/MRI- scanner [14]. A comprehensive review [3] illustrates the evolution of preclinical PET systems and their detectors, leading up to the current state-of-the-art with a specific aim of achieving high spatial resolution (SR).

Today's commercial high-resolution PET systems generally employ pixelated or monolithic scintillators coupled with silicon photomultipliers (SiPMs). The spatial resolution (SR) of pixelated scintillators is limited due to the size of their crystal cross-section (planar direction, perpendicular to the detector normal). For higher detector SR, researchers reduced the crystal cross-section to below 0.5 mm in small-animal PET scanners [15], [16]. However, this approach raises production costs and decreases detector efficiency and system sensitivity due to dead space caused by non-scintillating components like reflective foils between the crystal segments. Commercial high-resolution systems still employ crystal cross-sections of around $1 - 1.8$ mm [3], [17]. Moreover, traditional pixelated designs lack depth-of-interaction (DOI) information, which indicates the gamma interaction position along the third crystal dimension. DOI significantly enhances PET images by reducing parallax errors caused by photons entering the detector at oblique angles [18]. This is especially crucial for small-diameter PET scanners, such as preclinical or organ-dedicated scanners, and local detector modules at a small distance to the source to obtain a homogeneous spatial resolution across the field of view.

In response, alternative scintillator designs have emerged, such as monolithic scintillators that comprise an undivided scintillator block, distributing the optical light across multiple readout channels [19]–[21]. The resulting light patterns are location-specific, enabling the extraction of both planar position and DOI. Various 3D position estimation methods have been explored, including parametric techniques like the center of gravity (COG) [22], Cauchy modeling [23], nonlinear least-square modeling [24], and local linear embedding [25], as well as statistical methods such as k-nearest neighbors [26], maximum likelihood [27], [28], neural networks [29], [30], Voronoi diagrams [31], and Gradient Tree Boosting (GTB) [32], [33]. While statistical methods typically require calibration procedures with an external reference, they can account for deviations in detector behavior, e.g., due to defects in the scintillator. Furthermore, they offer improved detector performance along the edges of the monolith, where bias effects can arise due to light reflections and truncations [19], [34]. However, the calibration process can be complex, and the bias effect still occurs at all four planar and both DOI sides. Another challenge that monoliths face is the large crystal area and large readout region that leads to increased pile-up and dead time effects. This is especially relevant for small-diameter scanners and dedicated/local scanner modules with higher count-rate statistics. In addition, if individual channel readout is used, low photon density could not be read out due to trigger thresholds, which can affect the spatial, timing, and energy resolution. Furthermore, the resulting higher data rates require a correspondingly higher-performance processing chain including firmware and hardware.

[1] Department of Physics of Molecular Imaging Systems, Institute for Experimental Molecular Imaging, RWTH Aachen University, Aachen, Germany

[2] Hyperion Hybrid Imaging Systems GmbH, Aachen, Germany

[3] Physics Institute III B, RWTH Aachen University, Aachen, Germany

A promising approach to reduce the challenges of monoliths is semi-monolithic detectors [35]–[38]. Slab-based designs combine the advantages of both monolithic crystals and segmented detectors by segmenting the monolithic block in one dimension (planar-segmented direction) while retaining the other monolithic dimension (planar-monolithic direction). This mitigates readout drawbacks and limits the planar edge effect to the planar-monolithic dimension. DOI information is intrinsically preserved. In the pursuit of high-resolution detectors, segmentation can be in the range of 1 mm, similar to pixelated designs [39]–[46]. Researchers achieved detector spatial resolutions in the range of 1 − 2 mm FWHM along the planar-monolithic direction and of 2.5 − 3 mm in the DOI direction at an energy resolution of 14 − 15% which will be discussed in chapter V.

In this work, we present a finely segmented slab detector that provides high spatial resolution at least competitive with comparable designs by applying 3D machine-learning position estimation. Both the detector design and calibration routines are tailored for practical use in high-resolution PET applications. The scintillator comprises LYSO slabs of dimensions (24 × 10 × 1) mm³ that are arranged in two arrays with a pitch of 1.1 mm, each containing 44 slabs, coupled to an array of 12 × 12 readout channels with a pitch of 4 mm. This split design along the planar-monolithic direction allows for a higher photon density which not only offers an advantage in terms of energy and timing performance, but the more limited light distributions reduce the detector dead time as well as the data stream that must be processed. Combined with the calibration routines, these features shall specifically enable application in small-animal scanners and systems with local detector module concepts such as the HD-MetaPET project, as well as further organ-dedicated applications in consideration of the 10 mm crystal thickness.

## II. MATERIALS

### A. Radiation Detectors

The radiation detectors used in this work are based on LYSO scintillators coupled with an array of readout channels based on the DPC 3200-22 (Philips Digital Photon Counting, PDPC) [47], [48]. The scintillator topography of the detector under test consists of highly segmented slabs. As readout electronics, the Hyperion III PET and PET/MRI detector platform was used [49], [50]. The platform allows recording the complete raw sensor data stream onto long-term storage of a data acquisition computer for, e.g., later offline analysis or use in a detector calibration process.

### 1) Photosensor

An array of 12 × 12 readout channels make up the photosensor array with 3200 single photon avalanche diodes (SPADs) per readout channel. With a readout channel pitch of 4 mm, the photosensor has a sensitive area of (48 × 48) mm². 2 × 2 readout channels are housed on a DPC3200-22 which is self-triggering and individually operating. If one readout channel passes the two-level trigger scheme (trigger and validation), all four readout channels on the DPC are digitized. Both threshold levels are configurable. In this work, trigger scheme 2 was utilized which statistically requires 2.33 ± 0.67 SPADs breakdowns [51] together with validation scheme 4 which has an average threshold of 17.0 ± 6.2 discharges to start the acquisition sequence. In case of a validated trigger and after the acquisition sequence, 4 photon counts, and one timestamp is returned for this DPC, which is called a hit. Those combined hits must be clustered in pre-processing steps to obtain a so-called cluster that contains the recorded information on a gamma interaction. A consequence of this architecture can be missing DPC readouts in the clusters, due to trigger probabilities and possible dead times. This may result in a different set of read-out DPCs for 2 gamma photons that interacted at the very same position with the same energy deposit, which can be a problem, especially for light-sharing detectors. All developed processing steps need to handle clusters with missing hit information to sustain the detector's sensitivity. For all measurements, the photosensor was operated at 2.7 V overvoltage.

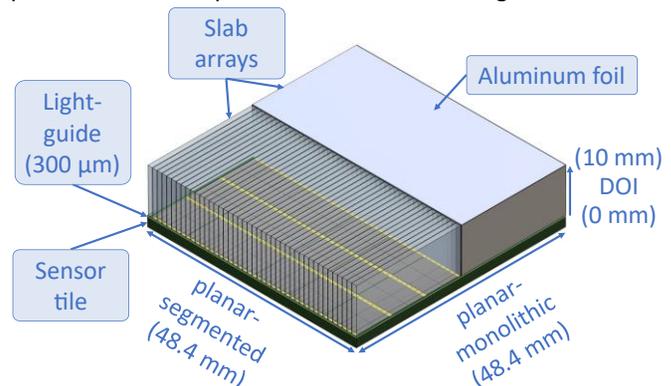

Figure 1: Schematic depiction of the semi-monolithic high-resolution detector design. Two slab array packages cover an array of 12 × 12 digital SiPMs based on 6 × 6 DPC 3200-22 (PDPC). A 0.1 mm BaSO4 layer surrounds the side surfaces of the slabs. A retroreflector was placed on top of each scintillator array. A layer of aluminum foil protects the outer crystal surfaces.

### 2) High-Resolution Slab Detector

The scintillator topology comprises high-resolution LYSO slabs of dimensions (24 × 10 × 1) mm³ arranged in two slab arrays with 44 slabs each and a 0.1 mm reflective layer of BaSO4 between and on the sides of the slabs (cf. Figure 1) which results in a total slab pitch of 1.1 mm. The fine slab segmentation requires light sharing across the planar-segmented direction to be able to resolve the single slabs on the photosensor with 4 mm readout channel pitch. This is ensured by a 0.3 mm light guide between the crystal and the photosensor. Both slab arrays grouped have a footprint of approximately (48.4 × 48.4) mm² which about matches the sensor tile's active area footprint of about (48 × 48) mm². A retroreflector was applied to the top surface of both crystal arrays. The top and side surfaces of each array were wrapped in aluminum foil for protection. The crystal array was glued to the photosensor using SCIONIX two-component optical interface glue [52].

### 3) Coincidence / Reference Detectors

For the flood irradiation setup (see section II.B.2)), a second high-resolution slab detector was used and for the collimator setup (see section II.B.1)) a one-to-one coupled detector. The one-to-one detector consisted of an array of $12 \times 12$ LYSO segments with 19 mm height and 4 mm pitch coupled to the same photosensor type.

### B. Coincidence Calibration Setups

**Planar configuration**

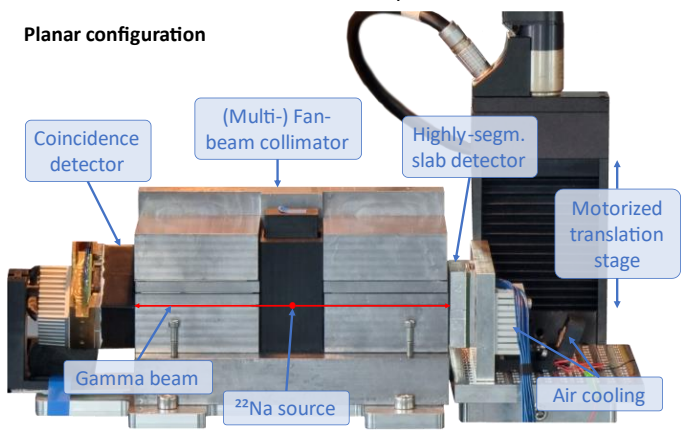

**DOI configuration**

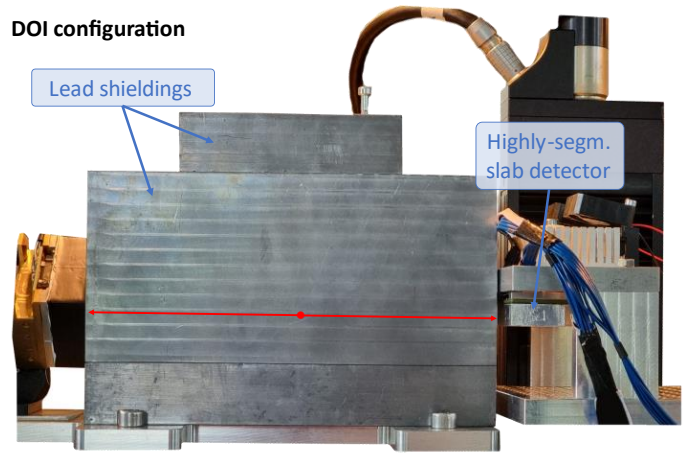

*Figure 2: Fan beam collimation setup similar to [32], [53], [54]. The radioactive point sources are housed in the center of the setup and confined to a fan beam by the surrounding lead plates. A detector under test and a coincidence detector were placed on both ends of the collimator with dedicated air cooling. The detector under test is mounted on a linear translation stage. Top: planar(-monolithic) configuration. For the planar-segmented calibration, the detector was rotated by 90° around the detector normal. Bottom: DOI configuration with the fan beam hitting the planar-segmented side of the detector. The detector is rotated by 180° to irradiate both slab arrays directly.*

### 1) Collimator Setup

For positioning calibration, a high-precision fan-beam-based coincidence setup was utilized [53], [54] with the detector under test mounted on a linear translation stage (LIMES 90, Owis, Staufen im Breisgau, Germany) with a nominal maximum positioning error of 2 μm (Figure 2). The setup was placed in a climate chamber at a stable measurement condition of 18°C ambient temperature with dedicated detector air cooling. The operating temperature of the detector was ~19.6°C for the planar irradiations and ~20.6°C for DOI irradiation. The translation stage allowed for detector irradiation at configurable positions, while the detector is mounted differently for the three dimensions of the detector: planar-segmented, planar-monolithic, and DOI (compare Figure 2). The fan beam was created by two lead plates at 0.5 mm distance that shaped the radiation of two $^{22}$Na point sources with 0.5 mm diameter and a total activity of about 18 MBq. The setup with the described coincidence detector generates roughly 600 cps of detected gamma interactions over the full width of the slab detector for the planar directions.

### 2) Flood Irradiation Setup

For energy and timing calibration, a second high-resolution slab detector was used as the coincidence detector in a flood irradiation configuration. Data was recorded for 9 source positions arranged in a $3 \times 3$ grid in the central plane between both slab detectors. The point source with a diameter of 0.5 mm and an activity of 2.1 MBq was moved with a motor stage. The operating temperature of both detectors was around 19.7°C.

## III. METHODS

### A. Data Acquisition and Pre-Processing

#### 1) General Pre-Processing Steps

The general pre-processing steps follow the scheme of clustering and coincidence search as described in [22] and apply to all used datasets. In the initial phase, the individual photosensor hits are grouped together into clusters within a specified cluster time window of 40 ns.

To filter clusters far outside of the uncalibrated photopeak at about 1650 optical photons for most slabs, only clusters with an optical photon sum exceeding 500 were considered for further analysis. More specific energy windows (EWs), that are relevant in practice, are applied to the datasets after an energy calibration, introduced in section III.D.

To reduce the computational cost of positioning all clusters, the clusters of the detector under test and the coincidence detector are compared with respect to their first timestamp. If the timestamps are within a coincidence window of 20 ns, both clusters form a so-called coincidence cluster.

#### 2) Calibration Setup Precision Evaluation

To evaluate the collimator setup's precision and the fan beam's actual expansion at the detector, we performed a coincidence count rate measurement according to [55] along the planar directions. A beam profile was obtained showing a full-width half maximum (FWHM) of $(0.48 \pm 0.01)$ mm. The planar crystal edges can also be estimated from the measurement, which, for example, is used to generate the motor stage grid for the positioning calibrations.

### 3) Positioning Calibration Datasets

The main aim of acquiring data for positioning calibration is to create labeled datasets that enable the training of supervised machine learning models designed for gamma interaction positioning. We developed separate machine-learning models for each direction of irradiation.

The original datasets were divided into distinct subsets for training, validation, and testing (60-20-20).

#### a)    Planar-Segmented Direction

To calibrate along the planar-segmented direction, the fan beam focuses on a single slab column (1 slab from each slab array) at a time, precisely at the center of each slab column. This approach results in a stepping pitch equivalent to the slab pitch of 1.1 mm, involving a total of 44 measurements. Each irradiation position took approximately 17.5 seconds to collect a minimum of 10,000 data points per irradiation position. Consequently, the total effective measurement time is kept under 13 minutes. For the supplementary comparative analytical method explained in section III.C.2), the training phase incorporated the flood dataset from section III.A.4), while the testing phase utilized the fan beam test dataset mentioned before.

#### b)    Planar-Monolithic Direction

The calibration process for the planar-monolithic direction involved aligning the fan beam to hit all the slabs simultaneously, with the fan beam being perpendicular to the sensor plane. The detector under test was stepped through the beam along its planar-monolithic direction with a pitch of 0.5 mm, excluding two central irradiation positions where both slab arrays would have been irradiated together. This resulted in a total of 95 measuring positions. Each position required approximately 37.5 s to gather a minimum of 20,000 data points per irradiation position, resulting in a total measurement time of less than 1 h.

The collected dataset was divided into two parts based on the arrangement of the slab arrays, as motivated in section III.C. Clusters were assigned to a specific slab array based on the fan beam irradiation position. This assignment must match the slab array containing the highest optical photon sum of the underlying readout channels.

#### c)    DOI Direction

For the DOI calibration, the detector was aligned with the fan beam parallel to the sensor plane, successively targeting both the planar-segmented side of the scintillator due to the large extent in the planar-monolithic direction. Using a stepping pitch of 0.25 mm, this setup generated 40 distinct labels. While an alternative DOI calibration approach involving angular irradiation [53] could have reduced the DOI calibration efforts, we opted for lateral irradiation as the initial step in assessing this detector concept.

Every irradiation orientation generates a dataset for the respective slab array that was hit first with the clusters filtered using the planar-monolithic position estimation. This filtering aimed to minimize geometric and scatter-related fan beam broadening effects. Additionally, the cluster count was normalized along the planar-monolithic direction to mitigate potential biases in the subsequent machine-learning model training phase. Afterward, the datasets from both orientations were merged.

To achieve a count of 30,000 coincident clusters per label, an effective measurement time of 24 seconds per orientation was sufficient, resulting in a total effective data acquisition time of 2 × 16 min.

### 4) Energy and Timing Calibration Dataset

The flood irradiation setup provided the dataset employed for both energy and timing calibration with approximately 6.5 million coincident clusters, gathered in less than one hour.

All clusters were positioned in all three dimensions using the positioning models. Furthermore, for timing calibration, the cluster energies derived from the prior energy calibration were incorporated. The dataset was subsequently divided into training and test subsets, maintaining a split ratio of 95% for training and 5% for testing purposes. Only the central irradiation position was used in the test dataset for CTR estimation. Energy windows of 50% and 15% were applied to the timing test dataset.

## B.  Slab-based Readout Characteristics

The number of DPCs read out is examined individually per slab based on the planar-segmented training dataset.

## C.  Gradient Tree Boosting Position Estimation

Utilizing the GTB machine learning technique as detailed in references [32], [33], we created independent positioning models for three directional components: planar-segmented, planar-monolithic, and DOI. These models establish relations between the light distribution, along with other computed features, with the known fan-beam irradiation position as the label of reference.

GTB is an ensemble learning algorithm, leveraging the advantages of decision trees and boosting. The methodology involves iterative training of a sequence of decision trees, where each subsequent tree corrects the residuals introduced by its previous trees. The final position estimation results from the cumulative outputs of all the trees. The GTB algorithm is characterized by several hyperparameters, including:

1.  *Number of decision trees*: GTB builds an ensemble of binary decision trees, and the number of trees determines the size of the ensemble. Typically, a larger number of trees can lead to better performance, but it also increases computational costs.

2.  *Maximum depth*: The maximum depth $d$ of a decision tree refers to the maximum number of decisions in a single tree.

3.  *Learning rate*: The learning rate adjusts the impact of the individual decision trees in the ensemble during training. A smaller learning rate can help to prevent overfitting while requiring a larger number of decision trees for the best performance.

4. *Input features*: Prior research has demonstrated that incorporating pre-calculated features help to derive the information related to the position from the individual channels faster and result in smaller GTB model sizes for a certain target performance. The following input features were used for all positioning models with further features for the planar-monolithic and the DOI directions based on the previous position estimations: raw photon counts of all 144 readout channels of the triggered DPCs per cluster, first moment (COG) and second moment of the light distribution computed from all available channels, readout channel with the highest optical photon count, DPC containing the highest optical photon count, photon sum projections, total photon sum, squared channel intensities on the main DPC, hottest channel normalized to the sum on that DPC, planar-segmented position estimation (for planar-monolithic and DOI model training), planar-monolithic position estimation (for DOI model training).

Deeper trees can capture more complex relationships in the data, however, the model complexity, and thus the memory requirement (MR), scales with the maximum depth d estimated by eq. (1) for a single decision tree.

$$MR\ (d) = (2^d - 1) \times 11B + 2^d \times 6B \qquad (1)$$

A hyperparameter grid search was conducted according to Table 1.

*Table 1: GTB hyperparameters for each position estimation used in a grid search. The number of decision trees states the maximum number. The training was stopped earlier when the RMSE did not improve for 10 rounds. The best determined hyperparameter is stated in round brackets.*

|  | Number of decision trees (best) | Maximum depths (best) | Learning rates (best) |
|---|---|---|---|
| *Planar-segmented* | 50 (50) | {4, 8, 10, 12} (10) | {0.05, 0.1, 0.2, 0.4, 0.7} (0.05) |
| *Planar-monolithic* | 1000 (831) | {4, 8, 10, 12} (10) | {0.05, 0.1, 0.2, 0.4, 0.7} (0.05) |
| *DOI* | 2000 (2000) | {4, 8, 10, 12} (10) | {0.05, 0.1, 0.2, 0.4, 0.7} (0.1) |

### 1) Positioning Performance Parameters

The positioning performance was evaluated based on the positioning error distribution (estimated position $z_i$ − irradiation position $Z_j$), also called point spread function (PSF). This includes SR defined by the FWHM (determined with a fitting routine based on the NEMA NU 4-2008 procedure for PET scanner performance characterization [56]), mean absolute error (MAE) as per eq. 2, bias vector from eq. 3, root mean squared error (RMSE) as per eq. 4, and the fraction of clusters predicted within X mm of the label denoted as $\text{Score}_{X\,mm}$. For the planar-segmented positioning, $\text{Score}_{0.5\,mm}$ corresponds to a correct slab assignment, and $\text{Score}_{1.5\,mm}$ that also includes clusters that are positioned in one of the neighbors of the slab being irradiated centrally.

$$MAE(Z_j) = \frac{1}{N}\sum_i^N |z_i - Z_j| \qquad (2)$$

$$Bias(Z_j) = \frac{1}{N}\sum_i^N (z_i - Z_j) \qquad (3)$$

$$\text{RMSE}(Z_j) = \sqrt{\frac{1}{N}\sum_i^N (z_i - Z_j)^2} \qquad (4)$$

$N$ represents the number of clusters. The performance parameters can be calculated for every irradiation position $Z_j$ or as a global parameter. The RMSE was used as the GTB loss function in this work.

Statistical errors on the MAE and SR performance parameters were estimated by determining the respective performance parameters for 10 subsets of the test datasets. The standard error of the mean was less than 0.01/0.03 mm for the MAE/SR in agreement with previous estimates.

### 2) Planar-Segmented Position Estimation

The objective of planar-segmented position estimation is to allocate the correct slab column of interaction position, resulting in 44 distinct labels for the GTB classifier. The best hyperparameters were obtained from a grid search with the parameters as stated in Table 1.

In a comparative approach, an analytical method for slab assignment was also implemented. This method relied on a peak finder applied to a 1D projection of the COG distribution, which is computed from raw optical photon counts along the planar-segmented direction. The identified peaks in the distribution represent specific slab positions, allowing to assign clusters to slab positions based on their proximity to the peak centers. This fitting was performed individually per slab array with a dataset separation at the central COG position. Although enhancements in separability are possible through parameters like photon sum or second moment, this simple COG routine serves as a fundamental method to gauge the performance of the GTB classifier.

### 3) Planar-Monolithic Position Estimation

Due to the continuous planar-monolithic direction with a corresponding labeled dataset, a GTB regressor is used in this model training. A special feature of the detector architecture used here is that the slabs are divided into two parts. This leads to a discontinuity in the bias vector at the transition between the two slab arrays due to the well-known bias effect towards the scintillator edges. Likewise, the RMSE as a loss function does not achieve a good positioning performance at this point. This physical artifact would be attempted by the model to be learned and optimized rather than learning the true underlying patterns, which would additionally lead to poor generalization to new data. A test showed exactly this behavior, so a separate model was trained per slab array. A hyperparameter grid search was conducted according to Table 1.

### 4) DOI Position Estimation

A GTB regressor model was trained for the DOI direction. A grid search was executed according to Table 1.

### 5) Memory-Requirement Analysis

Memory requirements were evaluated for the GTB models with the hyperparameters as reported in Table 1 using the concept of Pareto efficiency. Specifically, concerning positioning performance, Pareto efficiency characterizes the scenario where enhancements in positioning performance can't be achieved without a corresponding increase in memory requirements, tied to GTB model complexity as per the earlier referenced eq. 1. Employing this framework, our analysis extends to the planar-monolithic and DOI directions, encompassing various configurations involving different numbers of trees, maximum depths, and learning rates.

### D. Energy Calibration

The detector crystal volume was partitioned into 44 × 42 × 4 discrete sub-volumes, called voxels. This voxelization yielding the best energy resolution was part of a grid search. All clusters were sorted into these voxels according to their estimated 3D positions. Each voxel was assigned a stable readout region, dynamically determined based on the mean light distribution of the respective clusters. Only readout channels contributing more than 1% to the mean light distribution were taken into consideration. During calibration, all clusters with a full read-out readout region were selected. The energy resolution was determined by fitting a Gaussian to the photopeak of an unseen test dataset after energy calculation per voxel. The fit range was $\pm 1\sigma$ around the photopeak. To address missing readout channels in the test dataset, the $k$-Nearest Neighbors ($k$NN) imputation technique [57] was employed. A total of $k = 20$ neighbors was chosen for this imputation based on clusters with a full readout. The uncertainty of the energy resolution was estimated by the standard error of the mean of 10 subsets of the test dataset.

### E. Timing Calibration

A convex timing calibration [58] was performed estimating time skews of different origins, e.g., electrical, and optical skews, which affect the individual channel time stamps. In a first stage, a matrix equation was set up correlating all different DPC combinations with their mean timestamp difference. The timestamps are based on DPC containing the highest optical photon count. By solving the resulting matrix equation, the DPC time-skews can be determined and corrected for. In a second stage, crystal-voxel-skews were corrected by dividing the scintillator volume into 4 × 4 × 4 voxels that results in four DOI layers. The matrix equation was filled with the time differences based on the first timestamp of each cluster. With the obtained time skews from this matrix equation, the final CTR can be estimated as the FWHM (Gauss fit) of the corrected time differences in the timing test dataset. Timing resolution uncertainty was determined using the standard error of the mean of 10 subsets of the test dataset.

## IV. RESULTS

### A. Readout Characteristics

Due to the irregular pitch of the slab array with respect to the sensor array (1.1 mm slab pitch compared to 4 mm readout channel pitch), some slabs, for example, are centered on a DPC column, while others are located between two DPC columns. This results in a different number of DPCs read out for the individual slabs. Figure 3 breaks down this distribution. There are 6 areas where mainly 3 DPCs were read out and 5 areas where mainly 6 DPCs were read out. This corresponds to the 6 DPC columns, respectively the 5 transitions between the DPCs. Furthermore, in a few cases, 2 DPCs or more than 6 DPCs were read out.

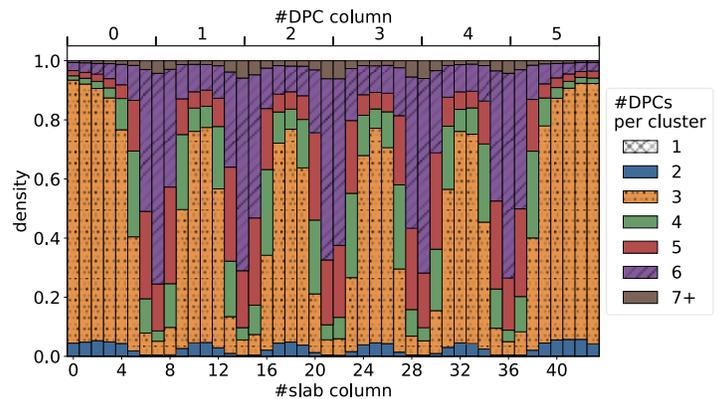

*Figure 3: Normed distribution of the number of read-out DPCs per slab. Densities below 0.5% were combined for illustration purposes. The DPC column positions are marked for comparison on top of the image.*

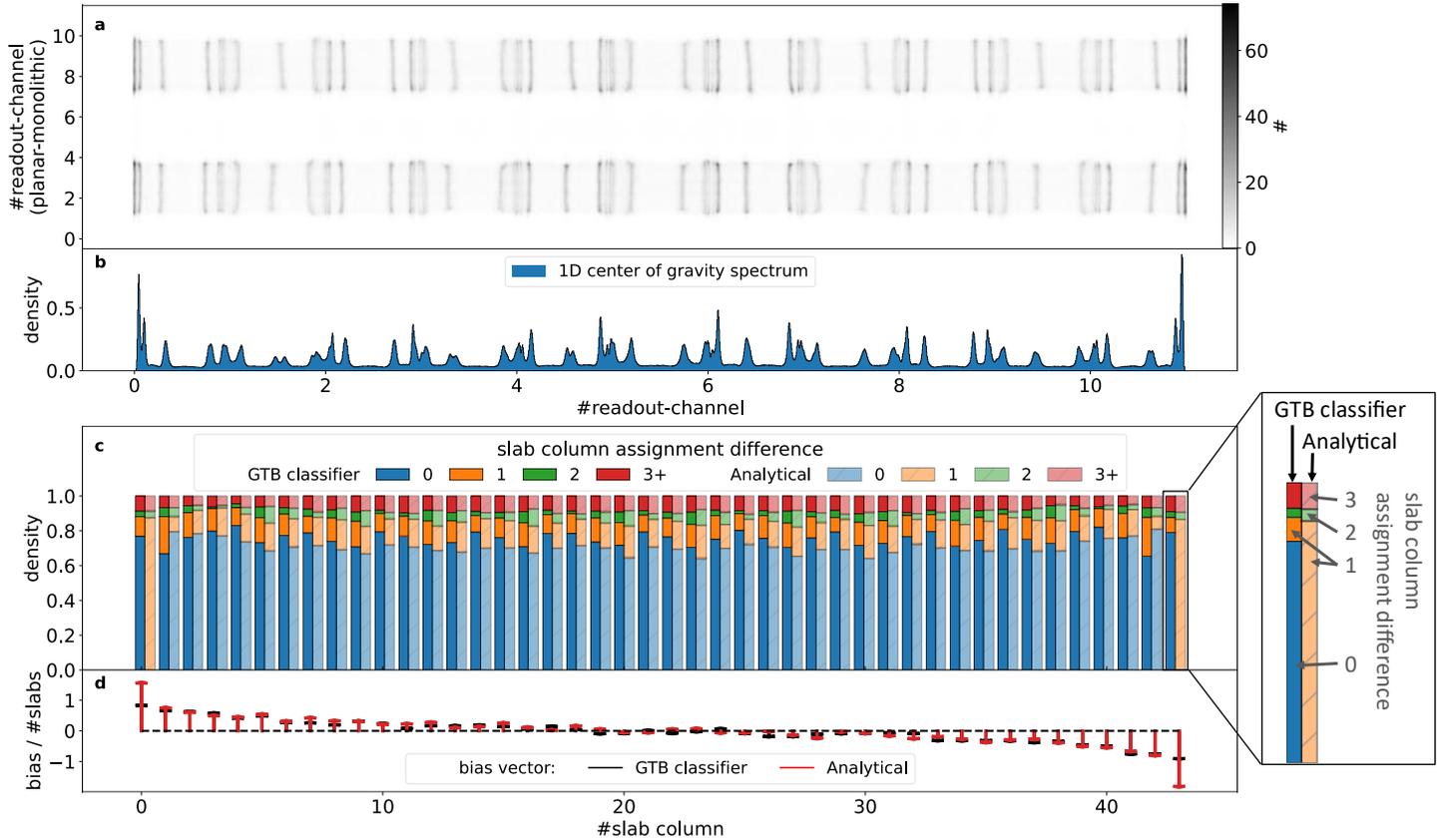

*Figure 4: Planar(-segmented) positioning and evaluation. (a) COG-based flood map without energy window showing the individual slabs as stripes in the planar-monolithic direction. The slab arrays and individual slabs are generally separable with a more pronounced overlay toward the slab array edges in the planar-segmented direction. (b) 1D projection of the COG distribution showing peaks that relate to the slabs. Individual peaks are separable, but all peaks are at least partly overlapping with their neighbors. (c) Slab column assignment difference per slab column at an energy window of (255-767) keV applied to the test dataset. A difference of 0 corresponds to a correct slab assignment of the clusters. A difference of 1 allows an assignment to a directly neighboring slab column. The GTB classifier is compared side-by-side with the analytical routine. (d) Bias vector of the slab column assignment difference per slab column.*

*Table 2: Positioning performance results for all three crystal dimensions with a 15 % and 50 % EW applied. $Score_{0.5mm}$ corresponds for the planar-segmented direction to a correct slab assignment, while $Score_{1.5mm}$ additionally counts the directly neighboring slabs as the correct assignment.*

|  | EW | MAE / mm | $Score_{0.5mm}$ | $Score_{1.5mm}$ | SR / mm |
|---|---|---|---|---|---|
| **Planar-segmented GTB (Analytical)** | 50% | - | 76% (67%) | 88% (86%) | - |
|  | 15% | - | 80% (74%) | 92% (92%) | - |
| **Planar-monolithic** | 50% | 0.93 | 50% | 84% | 1.27 |
|  | 15% | 0.75 | 55% | 88% | 1.18 |
| **DOI** | 50% | 1.13 | 33% | 74% | 2.35 |
|  | 15% | 1.03 | 36% | 78% | 2.14 |

## A. Positioning Performance

### 1) Planar-Segmented Direction

Figure 4 a shows the 2D COG distribution of the detector block. The dark stripes represent the slabs that are grouped in the two slab arrays that are identifiable as two rows of vertical stripes. Deviations between both slab arrays are evident that lead to, e.g., double peaks (compare the seventh stripe at #readout-channel ~ 1.7) in the 1D projection of the 2D COG distribution (subplot b). Labeled fan beam data with each slab irradiated at a time showed that the two outermost slab columns on each edge are overlapping in the COG distribution. Therefore, in the analytical peak fitting routine, all clusters in the outermost slabs were assigned to the second outermost slabs. This results in a slab column assignment difference of 1 or more (subplot c) that is also reflective in the bias vector (subplot d). The third outermost peak required a dedicated peak finder iteration to be resolved consistently.

The GTB classifier performed best with the hyperparameters: number of trees = 50, depth = 10, and learning rate = 0.05. The outermost slabs could be separated with a correct identification (slab column assignment difference of 0) of about 75% for the (255-767) keV EW. Overall, the GTB classifier had a higher correct slab identification for all slab positions compared to the analytical method except for the second and third outermost positions due to the not-identified second outermost slab. This also applies to the for $Score_{1.5\,mm}$ (slab column assignment difference of ≤ 1). Averaging all slab position assignments at the EW of (255-767) keV, the GTB classifier achieved a correct slab column identification of 76% (compare Table 2) while the analytical method achieved 67%. Considering a slab assignment in the irradiated slab or the directly adjacent ones, the GTB classifier scores 88% and the analytical method 86%. For the 15% EW, the correct and directly adjacent assignments score 80% and 92%, respectively, for the GTB classifier and score 74% and 92% for the analytical

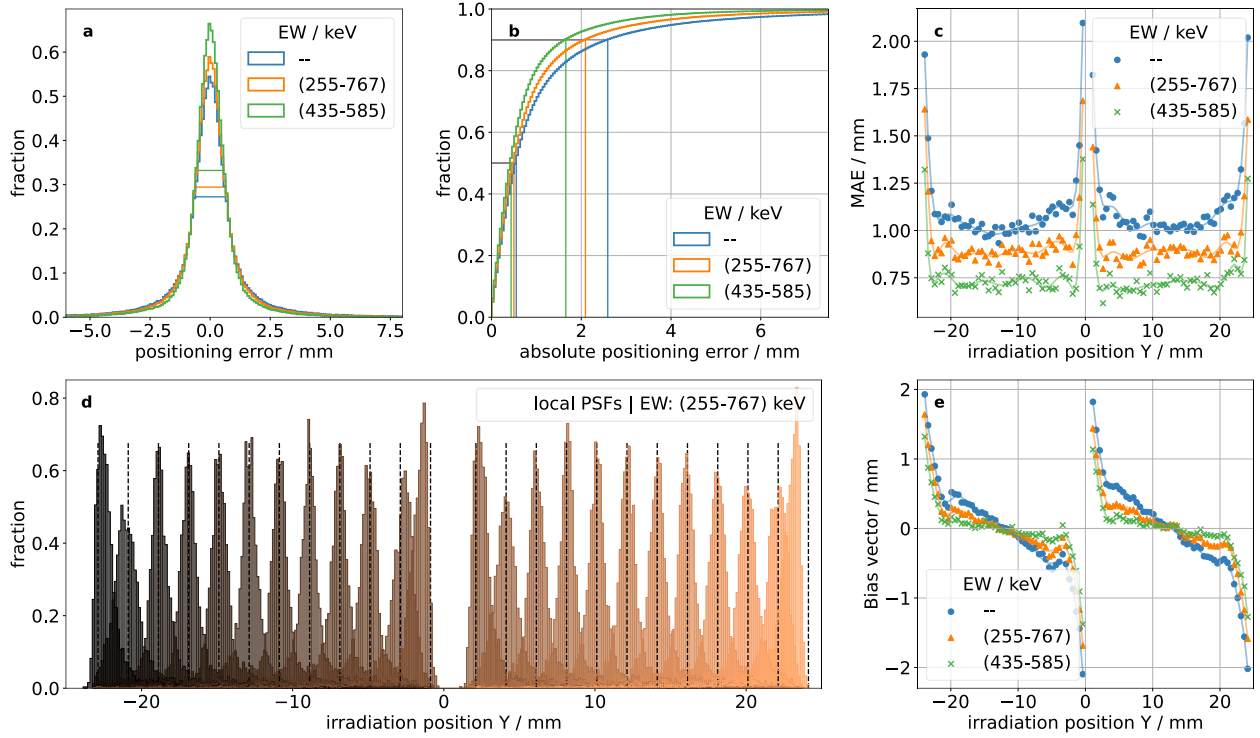

*Figure 5: Positioning parameters for the planar-monolithic direction at three energy windows. Subplot (a) shows the global PSF distributions and their cumulative illustration in subplot (b) that also includes the $50^{th}$ and $90^{th}$ percentile drawn as lines. The local MAE is depicted in subplot (c) for both slab arrays, and the local bias vector is shown in subplot (e). The transition between both slab arrays is approximately at the irradiation position of 0 mm. Subplot (d) shows the PSF distributions for different irradiation positions that are marked by vertical dotted lines. The energy window is (255-767) keV.*

method.

### 2) Planar-Monolithic Direction

Figure 5 and Table 2 summarize the best positioning performance results for the planar-monolithic direction achieved with a number of trees of 831, a maximum depth of 10, and a learning rate of 0.05. For the EW of 50/15%, a MAE of 0.93/0.75 mm is obtained, and a SR of 1.27/1.18 mm FWHM. The bias effect remained confined to a maximum of 2 mm, dissipating within a span of 2 – 3 mm from the edges (EW: 435 − 585 keV), despite the discontinuity between the two slab arrays. Figure 5d depicts the PSF locally for irradiation ranges of 2 mm marked by the dotted lines. For the 50% EW, the local PSF distributions follow the MAE and bias vector distributions with broader PSFs toward the crystal edges and a shift of the mean toward the crystal center of the respective slab array.

### 3) DOI Direction

Figure 6 and Table 2 display the performance of the DOI positioning for the best-performing hyperparameters of 2000 trees, a maximum depth of 10, and a learning rate of 0.1. The distribution of the PSF demonstrates a reduction in tails for narrower EWs. The local MAE and bias vector distribution affirm this trend across all irradiation positions. A bias effect is observed at both edges of the crystal along the DOI direction. This bias effect is more pronounced for DOI positions near the top crystal surface (DOI = 10 mm) compared to positions closer to the sensor tile (DOI = 0 mm). This disparity is also evident in subplot (d) of the local PSF distribution, where the mean of the highest DOI position range noticeably shifts inward within the crystal volume. For smaller energy windows, the bias effect at the edges is reduced, the saddle point remains at an offset bias of about 0.3 mm.

### 4) Memory-Requirement Optimization

The planar-monolithic model of one slab array shows a MR in the range of 10 MB and the DOI model around 30 MB. The Pareto front in Figure 9, based on a respective grid search, shows that the MR can be noticeably reduced with only minor losses in the MAE positioning performance. For the planar-monolithic positioning, at 99% MAE performance, the MR can be reduced by a factor of 10, at 95% performance by a factor of 250, and to a model size of 4 kB with 90% MAE performance. The DOI positioning shows a similar picture with a MR reduction by a factor of 750 to a model size of 40kB at 90% MAE performance.

### B. Energy Resolution

The found best voxelization divided the crystal volume into the slabs and in both monolithic directions roughly to their respective spatial resolutions. In Figure 7, showing the number of optical photons in the photopeak, six regions can be identified along the planar-segmented x-direction, separated by columns of fewer optical photons in the photopeak, that match the photosensor layout. Furthermore, in the outermost slabs, a noticeably lower number of optical photons is detected in the photopeak. With the voxel-based energy calibration, an energy resolution of $(13.8 \pm 0.1)$ % was achieved for the energy test dataset, as shown in Figure 8.

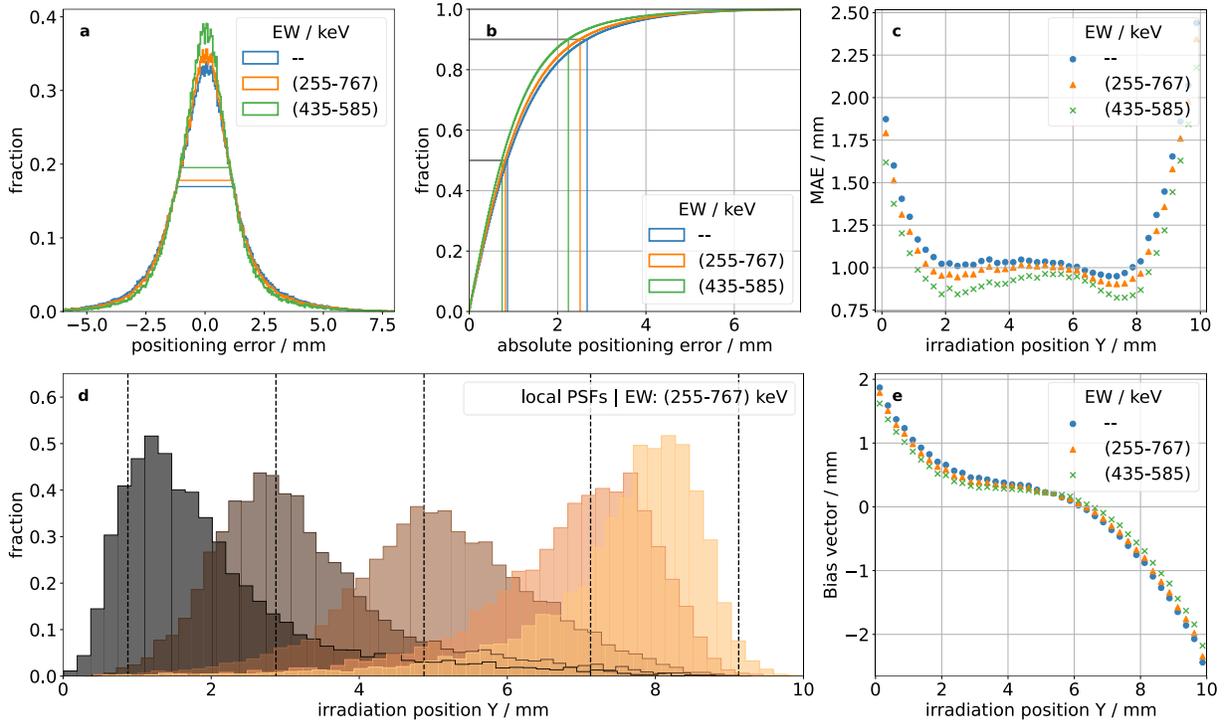

Figure 6: Positioning parameters for the DOI direction at three energy windows. Subplot (a) shows the global PSF distributions with a cumulative illustration in subplot (b) that also includes the 50th and 90th percentile drawn as horizontal lines. The local MAE is depicted in subplot (c), and the local bias vector is shown in subplot (e) with DOI 0 mm corresponding to the crystal surface close to the photosensor and DOI 10 mm corresponding to the top crystal surface. Subplot (d) shows the PSF distributions for five different irradiation positions marked by vertical dotted lines. The energy window was (255-767) keV.

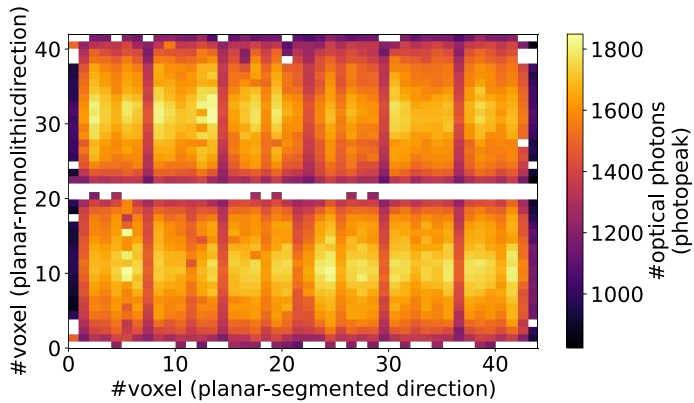

Figure 7: Photopeak mean position broken down into the planar voxels from the energy calibration. The DOI voxels were combined. Voxels colored in white were not calibrated due to, e.g., insufficient statistics. This mainly affects the edge regions of the slabs.

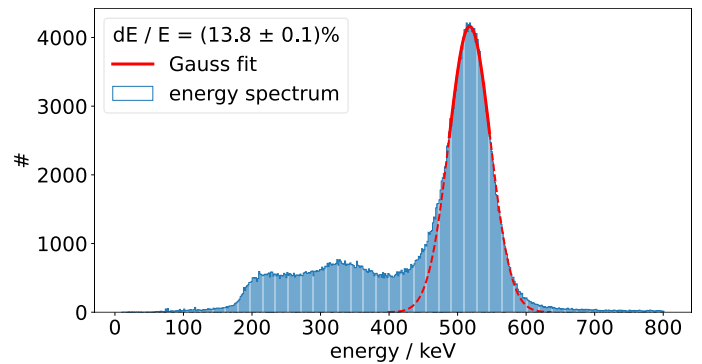

Figure 8: Calibrated energy distribution (blue) of the energy test dataset for all clusters. A Gauss fit was conducted in the range of ±1σ around the pre-fitted photopeak. The dashed red line shows a wider interval.

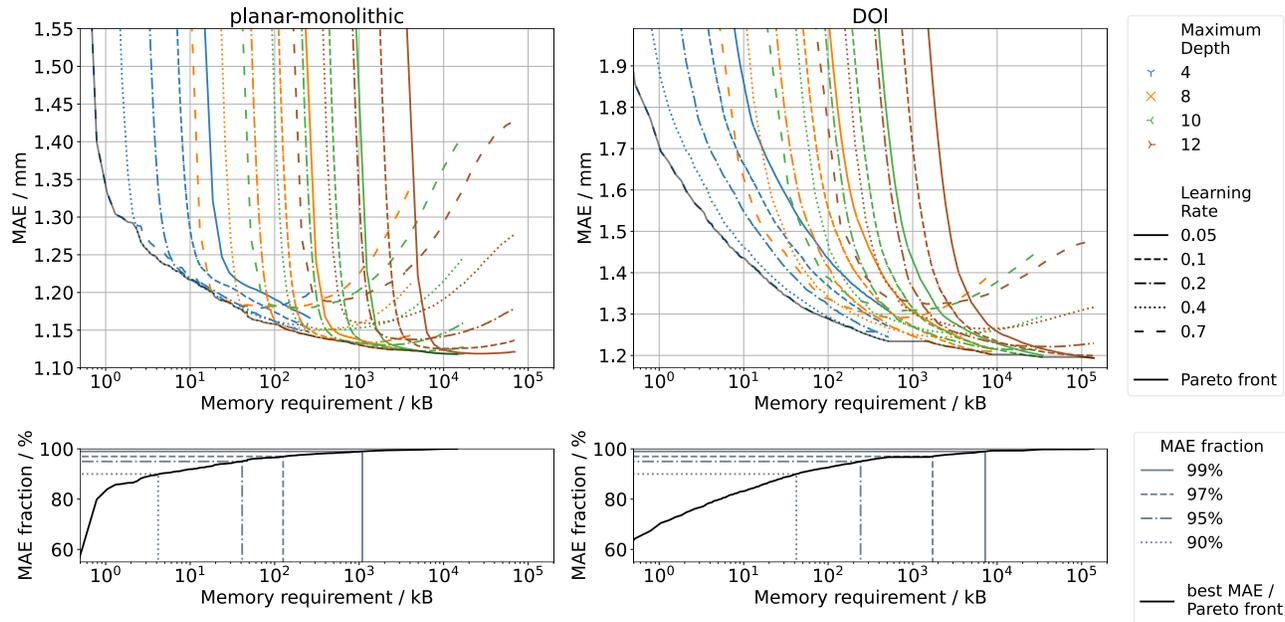

*Figure 9: The plots on the top depict the MAE performance across various learning rates and depths with respect to the MR in a logarithmic scale. The black Pareto front represents the optimal MAE for a specific MR. The left side displays the results for the planar-monolithic direction of one slab array, while the right side pertains to the DOI direction. The bottom subplots highlight the Pareto front normed by the best obtained MAE.*

## C. Coincidence Timing Resolution (CTR)

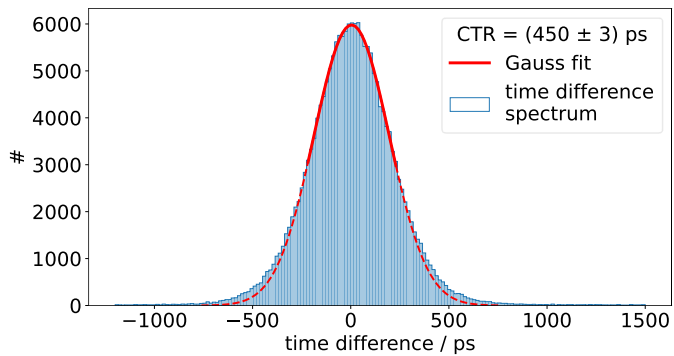

*Figure 10: Time difference spectrum (blue) of the timing test dataset at a 15% EW for both detectors. A Gaussian fitting was performed within the range of $\pm1.2\sigma$. The dashed red line illustrates a broader interval.*

For the EWs of 50 / 15% that apply to both detectors, a CTR of $(482 \pm 2)$ / $(450 \pm 3)$ ps FWHM was obtained. The Gauss fit matches the time difference spectrum in Figure 10 with a small deviation at the tails.

## V. DISCUSSION

The ability of the GTB classifier to identify slabs that were overlapping in the COG distribution indicates that the algorithm learns further patterns based on the provided input features. It was seen that the photon sum for the outermost slabs is lower, as well as slightly the squared channel intensities on the main DPC and the hottest channel normalized to the sum on that DPC. Incorporating these features into the analytical routine could improve this approach. When allowing directly adjacent slab assignments, both methods perform about equally which underlines that the classifier learns subtleties of other features, or combinations thereof, which allow a better slab identification compared to only using the first moment. This also includes the effect of missing DPC data. For the analytical method, it was, furthermore, important to distinguish between the two slab arrays. Otherwise, effects like a double peak for one slab column occurred due to geometric shifts. It must be noted that the analytical routine as well as the GTB classifies can be further optimized.

The two monolithic dimensions showed appealing positioning performance characteristics using a GTB regressor. With an SR 1.18 mm FWHM and a MAE of 0.75 mm in the planar-monolithic direction as well as 2.3 mm FWHM and 1.1 mm MAE for DOI at 15% EW, the detector design is feasible for high-resolution PET applications. This includes preclinical PET, such as small-animal imaging, and, keeping in mind the 10 mm crystal thickness, some organ-dedicated applications. Additionally, the design is an interesting candidate for evaluation within the HD-MetaPET project [14] as part of the high-resolution local detector module. The module will be placed within a LAFOV PET scanner for local spatial resolution enhancement.

Furthermore, the detector performs at least competitively to comparable architectures. A research group investigated similar slab dimensions of $(1 \times 11.6 \times 10)$ mm³ [40] and $(1 \times 37.6 \times 10)$ mm³ [46]. For the larger slab size, the group achieved 1.8 mm FWHM (planar-monolithic) and 2.5 mm FWHM (DOI) with improved performance through black surface treatment. Energy resolution was reported at 15.8%, and coincidence timing resolution (CTR) at 596 ps. This design was adapted for clinical PET scanners with larger slabs of dimension $(1.37 \times 51.2 \times 20)$ mm³ [41], yielding decreased performance parameters.

Another investigation with slabs of dimensions $(0.95 \times 25.4 \times 12)$ mm³, coupled with a 0.5 mm thick light guide to a photosensor featuring a 3.2 mm readout channel pitch, achieved an energy resolution of 13.7% and a CTR of 306 ps using the TOFPET2 ASIC with settings optimized for best CTR [42]. While the slabs were distinguishable in a flood map, no specific positioning performance of the detector was provided. A similar study reported 19% energy resolution at 281 ps CTR without specifying the measurement settings and filters applied [39]. Additionally, trapezoidal slab designs were examined for a compact ring setup, yielding high spatial resolution up to 1.05 mm FWHM in the planar-monolithic direction (after correcting for the source diameter) and sub-3 mm FWHM for DOI, accompanied by an energy resolution of around 14% [43]–[45]. CTR data were not provided.

For a comparison between detector performances, it is important to keep the underlying dataset preparation the same, for example in terms of EW and further filters. In this work, we have taken care to include all gamma interactions such as Compton and inter-crystal scatter to better fit PET system data. To focus on single-interaction photoelectric absorption, other groups introduce a COG filter/Anger mask around the fan beam irradiation position [38], [59]. By cutting away ±10% of the COG tails, we see a MAE improvement to 0.66 mm at 15% EW, while the SR describing the central area remains the same.

The bias vector and local PSF distribution indicate that the lower DOI layers are better separable than the upper layers. This behavior was also observed for wider and thicker slab designs [53]. The achieved spatial resolution, however, confirms the suitability of the detector for PET systems with small diameters, such as small-animal scanners.

Comparing the hyperparameter grid search for planar-monolithic and DOI positioning, the DOI model converges at a higher number of trees. This can be explained by the more complex dependency between DOI and the input features. An individual examination showed little DOI information on single input features.

Optimizing MR is critical to efficiently deploying machine learning models like GTB in high-throughput scenarios, such as efficient CPU implementation [60]. Furthermore, recent work has demonstrated direct integration into a Field Programmable Gate Array (FPGA) [61], which is inherently resource-constrained. The trained positioning models can be adapted accordingly with sizes up to a few kB under loss of below 10% MAE performance. The Pareto front converges for DOI slower compared to the planar-monolithic direction, resulting in a higher MR for the best MAE performance. The trade-off between model size and positioning performance depends on a case-by-case basis and can be determined individually per dimension. Additional MR optimizations could be achieved by reducing the number of input features, e.g., by cutting the 144 raw photon counts per cluster to a smaller readout region on the slab array or slab level. Further investigation is needed here to find the best balance between model variability, MR, and image quality.

The chosen detector design with split slab arrays and a light guide of 0.3 mm enabled light distributions over 3 or 6 DPCs, depending on the slab position. Therefore, the readout region and thus the data rate is reduced compared to the 36 DPCs on the photosensor, which allows for increased count rate statistics and less detector dead time. Due to missing DPC readouts, light sharing between the slab arrays, and Compton scatter, in some cases less than 3 or more than 6 DPCs contributed to a cluster.

Missing DPC readouts can significantly influence positioning and energy determination. GTB can handle these separately and the shown energy calculation corrects for this effect per imputation. Thus, the routines can also offer an advantage for other detector and readout configurations such as analog SiPMs digitized with an ASIC, e.g., the PETsys TOFPET2 ASIC [62]–[64].

The corresponding light densities allowed not only good spatial resolution but also an energy resolution of 13.8%, considering all clusters, which is competitive with similar detector designs as stated above and preclinical systems in general. Tests indicate that employing a thicker light guide negatively impacts energy resolution, likely attributed to the lower optical photon density and the readout channel thresholds. Consequently, this leads to a loss in optical photon statistics. Opting for a thinner light guide might introduce saturation effects owing to the constrained light distributions, potentially decreasing the separability of the slabs.

The outermost slabs show a lower number of optical photons in the respective photopeak due to optical photon losses towards the outside which is probably also reinforced by the slightly larger dimension of the slab array in the planar-segmented direction compared to the sensor tile's active SiPM area.

The time resolution of 450 ps is within the range of comparable designs and thus, like the other current detector designs, will rather have a small influence on the final image, depending on the application area [65]. However, noticeable CTR improvements can be anticipated by employing lower trigger thresholds (trigger scheme 1) [66] or state-of-the-art analog SIPM + ASIC digitization [39]. Lower thresholds come with certain trade-offs, including an increase in dead time, dark count rate, and heat generation. A machine-learning-based timing calibration showed significant improvements on a detector design with a 4 mm slab pitch [67], which remains to be analyzed on the finely segmented design.

## VI. Conclusions

The developed highly segmented slab detector shows promising characteristics for implementation in high-resolution PET scanners such as preclinical small-animal systems, local detector modules, and organ-dedicated applications. Using machine-learning-based 3D positioning, attractive positioning performance in all dimensions was achieved for 50% and 15% EW with up to 80% correct slab identification, 0.75 mm MAE along the planar-monolithic direction, and 1.03 mm MAE in DOI. The calibration routines are practical for PET scanners or dedicated PET modules with a smaller number of finely segmented slab detectors. Using scaled calibration methods, such as in-system positioning calibration with a fan-beam-based concept as shown in [53], applications in larger scanners are also conceivable.

## VII. Acknowledgement

This work was funded by the German Federal Ministry of Education and Research under contract number 13GW0621B within the funding program 'Recognizing and Treating Psychological and Neurological Illnesses - Potentials of Medical Technology for a Higher Quality of Life' ('Psychische und neurologische Erkrankungen erkennen und behandeln - Potenziale der Medizintechnik für eine höhere Lebensqualität nutzen').

We thank the scientific workshop of the RWTH University Hospital Aachen for manufacturing some of the technical components and Bruker for providing us with the scintillator material.

## VIII. Conflict of Interest Statement

The authors have no relevant conflicts of interest to disclose.